# Real-time detection of C-reactive protein in interstitial fluid using electrochemical impedance spectroscopy – towards wearable health monitoring

Aristea Grammoustianou[1], Ali Saeidi[1], Johan Longo[1], Felix Risch[2], Adrian M. Ionescu[1,2]

[1]Xsensio SA, EPFL Innovation Park, 1015 Lausanne, Switzerland
[2]Nanoelectronic Devices Laboratory, Ecole Polytechnique Fédérale de Lausanne, 1015 Lausanne, Switzerland

**Keywords**: Biosensing, Wearables, Electrochemical Impedance Spectroscopy (EIS), Single frequency impedance, C-reactive protein, Healthcare monitoring

**ABSTRACT**

Traditional detection methods of C-reactive protein (CRP) inflammation biomarker, in blood are expensive, time-consuming and labor-intensive. Such existing point-of-care CRP detection devices remain invasive, since they need blood sampling (finger-pricking or venous puncture). Here, we propose an electrochemical impedance spectroscopy (EIS)-based sensor for the real-time, fast, specific, sensitive, and label-free detection of C-reactive protein in the interstitial fluid (ISF) that can be accessed with minimally invasive microneedle arrays. The sensor has the potential to be integrated in a wearable device similar with the continuous glucose monitoring, that will detect CRP in interstitial fluid in a non-invasive, inexpensive and straightforward manner. The affinity-based assay was tested in both buffer and ISF-like solution. The limit of detection achieved was 0.7 µg/mL of CRP in buffer, and 0.8 µg/mL of CRP in ISF-like solution and the sensor shows excellent linearity up to 10 µg/mL. It is worth noting that the proposed sensor operates in low sample volume (down to 5 µL), and has a response time of 100 seconds.

**Keywords**: wearable sensor, interstitial fluid, electrochemical impedance spectroscopy, C-reactive protein, healthcare monitoring

## INTRODUCTION

C-reactive protein (CRP) is an acute-phase protein produced by the liver. It is recognized as a non-specific biomarker for cardiovascular disease and inflammation(Bryan et al., 2013). In general, the CRP level in plasma for a healthy adult is <3 mg/L, and it might increase 1000 times during acute phase of inflammation(Bryan et al., 2013), with bacterial infection being a powerful trigger for such an increase (Coelho et al., 2007). In addition, a number of studies have linked elevated CRP concentrations with an increased risk of cardiovascular diseases, Type 2 diabetes mellitus, age-related macular degeneration, hemorrhagic stroke, Alzheimer's disease and Parkinson's disease (Luan and Yao, 2018). It becomes clear that CRP is a biomarker of interest, the monitoring of which in the human plasma could facilitate the diagnosis and treatment of numerous conditions.

Laboratory-based turbidimetric and nephelometric technologies are the gold standard for CRP detection. Human CRP enzyme-linked immunosorbent assay (ELISA)-based techniques are also routinely used. Nevertheless, these techniques are expensive, labor-intensive, and require trained personnel. They are time-consuming and prone to false-positives; ELISA for example, requires labeling of antibodies/antigens and many incubation and washing steps. In the recent years, point-of-care (PoC) devices for CRP detection have emerged as attractive alternatives to the existing lab-based methods, thanks to their user-friendliness, portability, cost-effectiveness and fast results. The majority of these devices make use of turbidimetric, fluorescence, chemiluminescence, lateral flow assays, and sandwich immunoassays, detecting CRP concentrations in human blood (Pohanka, 2022), (Vashist et al., 2016). Having to sample blood, by finger-pricking or venous puncture, renders these PoC devices practically invasive and complex enough to be easily used by non-trained people. Interstitial fluid (ISF), is part of the extracellular fluid (ECF) between the cells (approximately 75% of ECF), and is considered an emerging blood alternate biofluid for tracking biomarker concentrations and dynamics, especially because it contains the main diagnostic biomarkers/analytes found in blood (Friedel et al., 2023). ISF is being increasingly studied for its utility in diagnostic biosensing devices, which would be minimally invasive, inexpensive and user-friendly. The present study combines for the first time the real-time electrochemical impedance spectroscopy (EIS) sensing technique with the detection of CRP in ISF. According to the literature, electrochemical biosensors have been considerably used for the detection of CRP. Nonetheless, the majority of these studies, involve more complex bioassays that make use of signal enhancement through nanoparticles (Ma et al., 2022), (Thangamuthu et al., 2018) or secondary antibodies (Jampasa et al., 2018), demand sophisticated chemical treatment of the sensing surface (Balayan et al., 2022) (Kanyong et al., 2020), (Gupta et al., 2014), and tend to bypass real-time monitoring of CRP detection. In the current work, emphasis is placed on the development of an affinity-based assay that will allow for the label-free, real-time, rapid, inexpensive, and straightforward detection of CRP in buffer, complex matrix and biofluids. The long-term aim of this work is to enable the development of a wearable device that will detect in a label-free manner the concentration of CRP in human ISF in real time.

EIS is a common electrochemical technique which detects a wide range of biochemical interactions over the sensing surface (DNA hybridization, protein-protein, and protein-DNA interactions) (Songjaroen et al., 2016). EIS-based sensing, typically involves a frequency sweep, followed by fitting a complex equivalent circuit model. This approach has drawbacks, including imprecision due to fitting errors and the inability to provide real-time monitoring of biomolecular interactions. This study introduces an innovative EIS application for real-time protein sensing. The proposed sensing system operates at a constant frequency while monitoring the electrochemical sensor's impedance. This method offers multiple benefits, including simplified near-real time data analysis, improved precision, and robustness; in addition it is highly miniaturizable.

**MATERIALS AND METHODS**

**1. Reagents**
Phosphate buffered saline (PBS) tablets, pH 7.4 (10 mM phosphate buffer, 2.7 mM potassium chloride and 137 mM sodium chloride) and bovine serum albumin (BSA) were purchased from Merck (Darmstadt, Germany). Neutravidin was purchased from Thermo Fisher Scientific (Massachusetts, US). Biotin anti-CRP antibody and biotin control non-CRP antibody were acquired from Abcam (Cambridge, UK). Human CRP and human CRP-depleted serum were purchased from BBI Solutions (Caerphilly, UK). All solutions used were prepared in PBS (pH 7.4), except from the solutions that were prepared in the ISF-like medium (human CRP depleted serum, diluted 2.5 times in PBS). Ethanol (EtOH) absolute 99.8% was bought by Fisher Scientific (Loughborough, UK). Milli-Q water (18 MΩ) was obtained using the Arium® Mini water system by Sartorius (Goettingen, Germany).

**2. EIS measurement setup**
EIS measurements were conducted using an SP-200 biologic potentiostat, where a DC voltage of 400mV and a low-frequency AC signal with a 20mV amplitude were applied. The AC signal's frequency was adjusted to a fixed value, tailored for each set of experiments to optimize the signal-to-noise ratio (SNR) based on the experimental configuration and biofluid characteristics (supplementary materials). The capacitive response, extracted from the imaginary part of the measured impedance ($Z_{Im}$) based on a simplified model, served as the sensing signal. Thin-film gold electrodes (ED-SE1-Au; Micrux Technologies, Asturias, Spain) were used as the sensing surfaces. Prior to the experiments, the electrodes were first rinsed with MilliQ (18 MΩ) water and 70% EtOH, and then subjected to UV ozon cleaning for 30 min (type F UV ozon cleaner, Ossila, Sheffield, UK). All measurements were taken at 25º C.

**3. Quartz crystal microbalance with dissipation (QCM-D) commercial device**
A QSense Analyzer (4-channel system) Biolin Scientific, Sweden) acoustic device was employed for real-time measurements of frequency changes. Gold (Au) coated 5 MHz quartz crystals (QSensors) were used. Frequency responses were recorded at 35 MHz overtone. Prior to the experiments, the QSensors were first rinsed with Milli-Q (18 MΩ) water and 70% EtOH and then subjected to UV ozon cleaning for 30 min (type F UV ozon cleaner, Ossila, Sheffield, UK). All measurements were taken at 25º C. The acoustic device was connected to a Gilson peristaltic pump - MINIPULS® 3 (Wisconsin, US), through which samples were injected into the sensing system.

   The QCM-D consists of an AT-cut quartz crystal sandwiched between two electrodes. Application of an alternating voltage through the QCM electrodes causes the crystal to oscillate at a fundamental resonant frequency, ranging between 5 and 50 MHz, resulting in the generation of a standing acoustic wave inside the crystal. The fundamental resonant frequency (F) of the crystal is sensitive to mass deposition on the crystal surface, according to Sauerbrey's equation (Sauerbrey, 1959).

**4. Sensor functionalization and label-free electrochemical detection of CRP**

The ED-SE1-Au sensing surface was coated with neutravidin dissolved in PBS (pH 7.4). Afterwards, the biotinylated anti-CRP antibody was surface-immobilized through neutravidin-biotin interaction. Subsequently, the sensor was passivated with BSA in PBS pH 7.4. Eventually, human CRP diluted in PBS (pH 7.4) at concentrations ranging from 0.7 to 10 µg/mL were flowed over the sensing surface. Each sample addition was followed by buffer washing. For the experiments carried out in the ISF-like solution, human CRP solutions were prepared in human CRP depleted serum that had been diluted 2.5 times (to match the total protein concentration of ISF (Haljamae H and Freden, 1970)). Again, the concentrations tested were ranging from 0.7 to 10 µg/mL. Unless otherwise stated, the volume of the human CRP samples injected in the sensing system was 200 µL and the experiments were done in a static-mode. All measurements were taken at 25ºC.

**5. Sensor functionalization and label-free acoustic detection of CRP**

The 5 MHz Au coated QSensor was modified following the experimental procedure described in section 4.

**6. Atomic Force Microscopy (AFM) imaging**

The surface modification was confirmed by AFM - topography imaging. A bare gold sensor was compared with a gold sensor coated with neutravidin, and a gold sensor coated with neutravidin and subsequently modified with the CRP antibody. Immediately after *in-situ* modification, the three samples were rinsed with MilliQ (18 MΩ) water and dried under a mild nitrogen gas stream. The experiments were performed in an air environment at room temperature with the standard S-Scanner of the Asylum Cypher system from Oxford Instruments (Abingdon, UK). Images (512x512 pixels) were taken in AC-tapping mode using Tap300Al-G tips from BudgetSensors® (Sofia, Bulgaria) at a scan rate of 1.95Hz.

## RESULTS AND DISCUSSION
### 1. Acoustic characterization of the CRP assay

Initially, the development of the CRP assay was validated with a commercial QCM-D acoustic device. QCM-D has been extensively used in the characterization of biological interactions on engineered surfaces (Tonda-Turo et al., 2018). Figure 1a summarizes the main surface functionalization steps of the CRP assay which were established and verified with the QCM-D acoustic device. For the CRP assay, the QSensor was coated with neutravidin that saturated the surface and enabled the oriented immobilization of the biotinylated CRP antibody. The surface modification steps were imaged with AFM. Figure 1c illustrates the topography of the QSensor after surface modification. Neutravidin adsorption on gold (fig. 1c-2) increased the roughness of the sensing surface as opposed to a bare gold sensor (fig. 1c-1). CRP antibody-neutravidin coupling further increased the roughness of the surface (fig. 1c-3). According to the QCM-D measurement, the frequency change upon neutravidin adsorption was 397 (±17.4) Hz (Fig. 2a). the Immobilization of the biotinylated CRP antibody caused a decrease in frequency of 219 (±7.93) Hz (Fig. 2b). The detection of 200 µg/mL of CRP by the specific CRP antibody led to a frequency decrease of 13 Hz (Fig. 2c). The specific concentration of 200 µg/mL was chosen to study the specificity of the assay. Hence, a control experiment was conducted, where all the assay steps remained the same, except from the CRP antibody immobilization which was replaced by the immobilization of the non-CRP control antibody at the same concentration. The non-CRP antibody could not detect CRP, as is shown in the inset of Figure 1c; no frequency shift was recorded by the addition of 200 µg/mL of CRP to the acoustic sensing system, proving that the bioassay could specifically detect CRP in PBS.

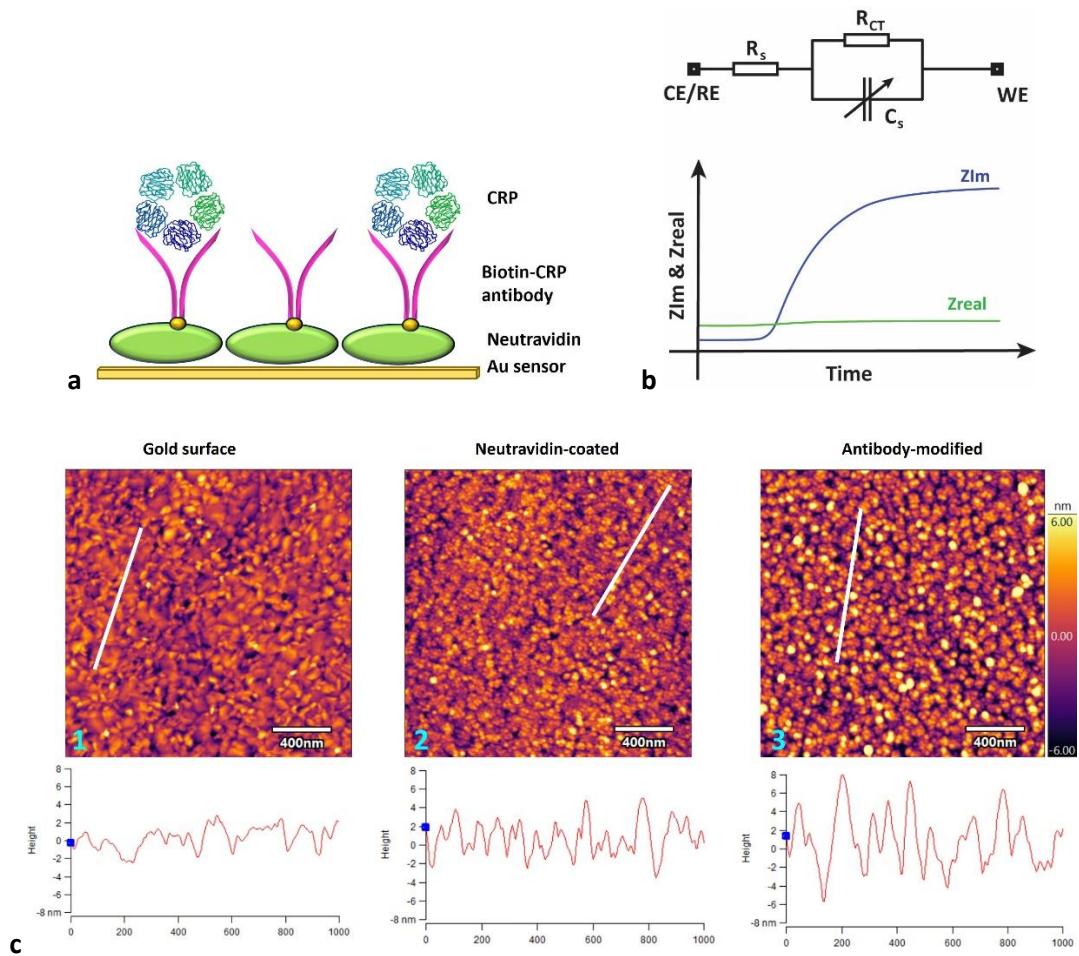

**Figure 1. a)** Schematic representation of the CRP assay. A gold (Au) surface of either an electrochemical or acoustic sensor was covered with a neutravidin layer, immobilized with the biotin-CRP antibody, and treated with CRP solution (not drawn to scale); **b)** Equivalent electrical circuit model of the sensor (top) and its expected response upon CRP detection (bottom); **c)** AFM topographic images and respective surface profiles portraying the surface functionalization steps: bare gold sensing surface (c1); gold surface coated with neutravidin (c2); gold surface coated with neutravidin and modified with the CRP antibody (c3).

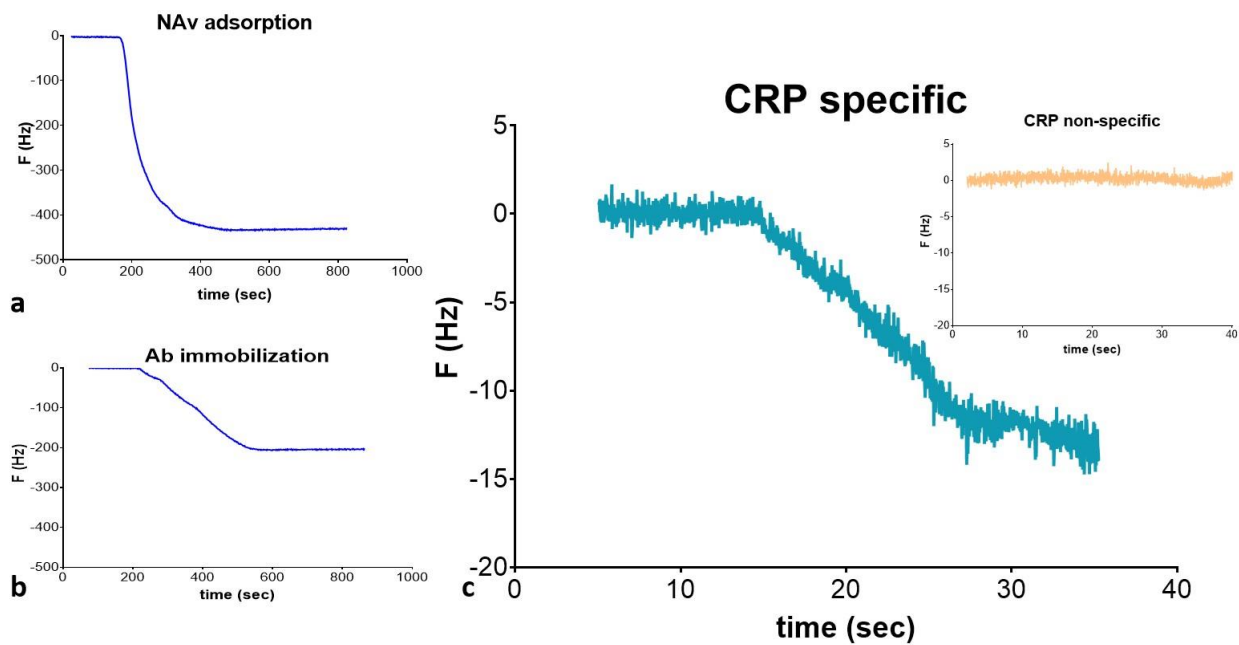

**Figure 2.** Real-time QCM-D graphs showing frequency signals upon: **a)** neutravidin (NAv) adsorption on the QSensor; **b)** CRP antibody (Ab) binding to neutravidin; **c)** CRP (200 μg/mL) detection. The inset depicts the absence of frequency shift in the case of the control experiment, where the CRP antibody was replaced by a control non-CRP antibody.

## 2. Transfer of the acoustic-based CRP assay to the EIS sensing setup

After the CRP assay was developed and validated on the commercial QCM-D device, it was transferred to the EIS sensing system. The EIS sensing setup is based on a 2-electrode electrochemical system comprising one working electrode (WE) and one shared counter/reference electrode (CE/RE), as illustrated in figure 1b. It operates by applying a continuous, constant-frequency alternating signal to the CE/RE and measuring the resulting AC signal at the WE. The discrepancy between the input signal and the measured signal allows for the real-time determination of the impedance of the electrochemical cell. It is expected that protein adsorption on the sensing surface alters the capacitance of the sensing electrode, and this change can be observed by monitoring the imaginary part of impedance (ZIm) (figure 1b). Further information on the measurement parameters and the optimization method is provided in the supplementary section. The surface functionalization procedure was the same with the acoustic surface functionalization and figure S1 shows the steps of the neutravidin adsorption on the gold EIS surface (Fig. S1a) and the subsequent CRP antibody immobilization (Fig. S1b). For example, neutravidin adsorption led to an impedance increase of 123 (±37.5) Ohm. The antibody immobilization resulted in a ZIm increase of ~ 800 Ohm after the buffer washing step. The real-time alteration in the imaginary part of electrochemical impedance (referred to as ZIm) exhibits a direct correlation with the frequency shift of the acoustic sensor, resulting from mass deposition on the sensing surface. In the case of an acoustic sensor, mass adsorption changes the sensor's fundamental resonant frequency. While, within an electrochemical cell, mass adsorption results in the modification of the sensor's capacitance, which can be observed as a change in the imaginary part of impedance.

## 3. Real-time EIS detection of CRP in buffer

Figure 3 presents the real-time EIS response upon CRP detection in PBS (pH 7.4). The limit of detection (LoD) achieved under the experimental parameters described in section 4 of experimental setup was 0.7 µg/mL of CRP (Fig. 3a, green line). The high resolution of the sensor allowed the distinction between the two close concentrations of 0.7 and 0.8 µg/mL of CRP (fig. 3a, dark blue line). The predominant source of the sensor's average 23-ohm noise level is primarily attributed to the electrical measurement setup and connections, as extensively detailed in the supplementary materials. However, the CRP sensor presented in this study exhibits a Signal-to-Noise Ratio (SNR) of 20, with normalization applied to the shift in ZIm for a concentration of 1 µg/ml. Figure 3b portrays the real-time EIS detection of higher CRP concentrations, namely 8 and 10 µg/mL. The higher the CRP concentrations detected, the bigger the ZIm shift.

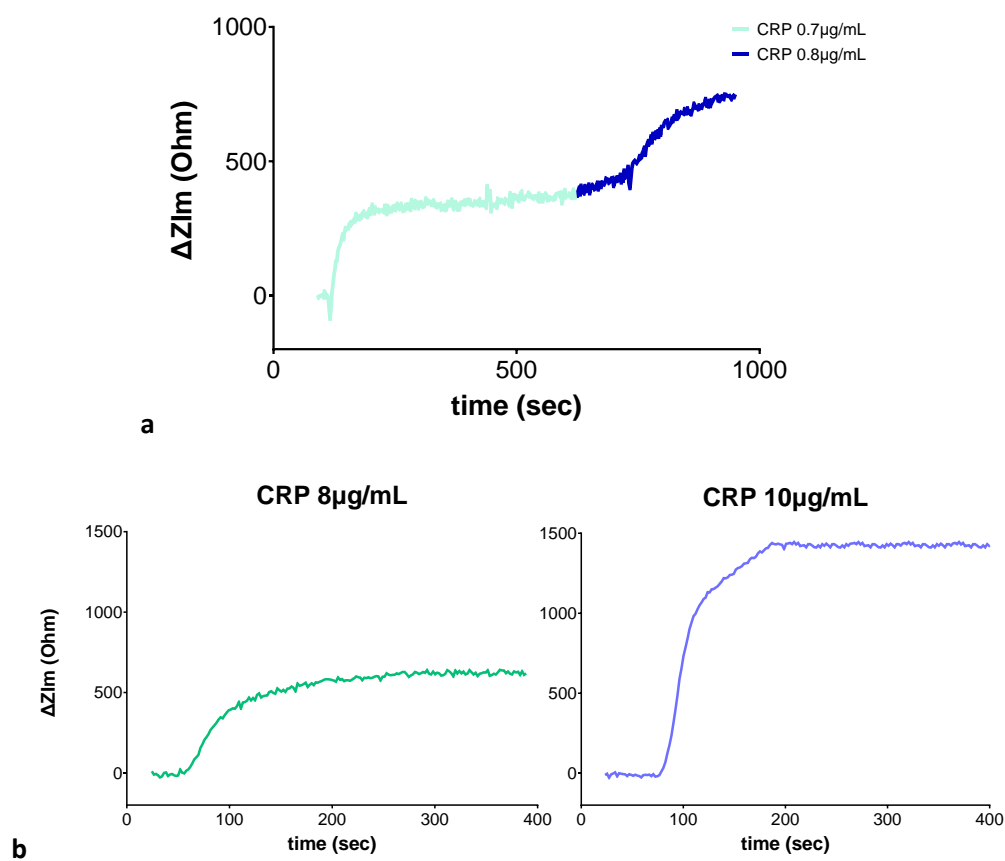

**Figure 3.** Real-time EIS detection of CRP in buffer: **a)** the limit of detection achieved was 0.7 µg/mL of CRP. At the same time, the sensor is able to distinguish between low concentrations of CRP, such as 0.7 µg/mL from 0.8 µg/mL of CRP; **b)** Sensor's response upon the detection of 8 and 10 µg/mL of CRP in buffer.

## 4. Detection of CRP in ISF-like medium

With the motive of devising a wearable sensor detecting CRP concentrations in ISF in a real-time, label-free and fast manner, the developed assay was tested against ISF-like solution. To obtain the ISF-like solution, human CRP depleted serum was diluted by 2.5 times in PBS, and then spiked with a range of CRP concentrations (0.7 to 10 µg/mL). To match the requirements of a wearable

ISF sensor, the experiments were carried out under a low flow rate: 2 µL/min. In short, the surface functionalization was done in a static mode, and then the sensor was transferred in a 3D printed flow cell (volume capacitance: 2 µL) for the actual CRP detection (Fig. S2). Before the addition of the CRP distinct concentrations, the sensor was conditioned with the CRP-depleted ISF-like solution. The absolute value of the ZIm shift was calculated at the time point of 100 sec after the CRP sample had reached the surface.

As shown in figure 4a, the limit of detection in the ISF-like medium increased to 0.8 µg/mL of CRP (ΔZIm = 60 Ohm), compared to the 0.7 µg/mL of CRP in buffer. Moreover, the magnitude of response was influenced by the ISF-like solution; comparison of the figures 3b and 4a, reveals a decrease in the ΔZIm of ~ 9 times for the same CRP concentration of 10 µg/mL (1420 Ohm in PBS versus 151 Ohm in ISF-like solution); simultaneously the noise level increased, resulting in the reduction of the SNR of the electrochemical measurement. This decrease is attributed to the complexity of the ISF-like matrix which interferes with the solid-liquid interface.

As can be seen in figure 4a, the ZIm response occurs when the CRP sample reaches the sensing surface (red arrows) and stabilizes within 100 sec, which is a characteristic that further meets the requirements of a real-time wearable ISF biosensor that is to be used in CRP detection.

The specificity of the assay was also investigated in the ISF-like medium. Figure 4b summarizes the sensor's specific and non-specific response upon the injection of 0.8 and 10 µg/mL of CRP in the sensing system. The response of the non-specific sensor (orange lines) which was modified with the control non-CRP antibody instead of the specific one, was subtracted from the specific response of the sensor (blue lines) and the resulting curves (black lines) were plotted. It can be noticed that for low CRP concentrations (close to the LoD), the signal-to-noise ratio is lower compared to high CRP concentrations, such as 10 µg/mL. An expected outcome, considering that when CRP is added in the sensing system at low concentrations, it competes with the high background signal considerably more than when it is added in the sensing system at higher concentrations.

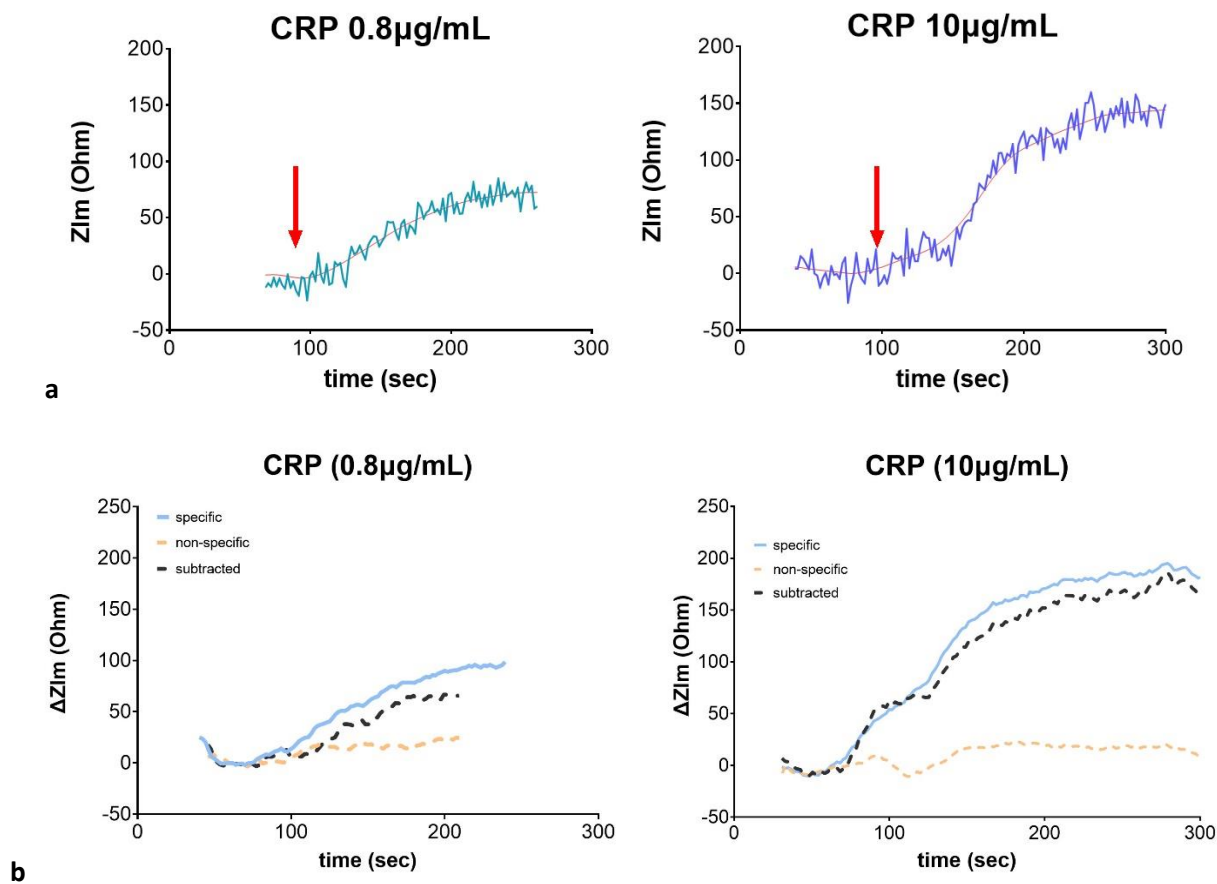

**Figure 4. a)** Real-time fitted graphs showing the detection of 0.8 μg/mL (60 Ohm) and 10 μg/mL (151 Ohm) of CRP in the ISF-like solution. The red lines represent the fitting; the red arrows indicate the time when the CRP samples reach the sensing surface; **b)** Subtraction of the non-specific from the specific ZIm response upon the injection of 0.8 and 10 μg/mL of CRP in the system. The blue lines show the specific response, the orange lines represent the non-specific response, and the black lines depict the subtraction.

## 5. Calibration curve and blind tests

After the specificity and sensitivity of the bioassay in ISF-like medium were determined, a calibration curve based on the ZIm changes upon CRP recognition in ISF-like solutions was performed. CRP concentrations ranged from 0.8 to 10 μg/mL. The sensor showed a linear response in the range of 0.8 to 10 μg/mL, with an $R^2$ of 0.9944 (Figure 5a).

To assess the robustness and reliability of the proposed CRP sensing platform, a blind test was performed. Four test samples were generated by spiking CRP in the ISF-like medium (2.5X diluted human serum), to obtain CRP final concentrations of 1 μg/ml, 3 μg/ml, 7 μg/ml, and 9 μg/ml. These samples were anonymized by an external individual, who assigned them neutral labels (referred to as sample 1 to 4). Subsequently, two samples, namely #1 and #3, were randomly picked up for the analysis and flowed through the sensing system, each, generating a ZIm response. The CRP sensing platform was employed to measure in real-time the ZIm shift, which was then correlated to the calibration curve to determine the concentration of the samples tested.

Figures 5b and 5c correspond to the ZIm real-time changes when each of the samples #1 and #3 were flowed over the surface. Sample #1 led to an ZIm increase of 130 Ohm, while sample #3 resulted in a smaller ZIm of 82 Ohm. Table 1 compares the actual CRP concentrations used to spike the tested samples with the CRP concentrations measured by the EIS sensor. The measurement error was calculated to be 7.7% for sample #1 and 3.4% for sample #3, respectively. An estimation of the error attributed to the sample preparation and handling was performed. CRP samples were sent to an external laboratory (Laboratoires, La Source, Lausanne, Switzerland), for concentration quantification using standard analytical devices (immunoturbidimetric sensing technique). More particularly, 2 sets of CRP samples of known concentrations, 5 replicates per set, were analyzed. One set represented a low CRP concentration in ISF-like medium and the other a high CRP concentration. Each set was analyzed in duplicate (using the Alinity C Series SCM02060 device; Abbott, IL, US) and the average error was found to be 2.3%.

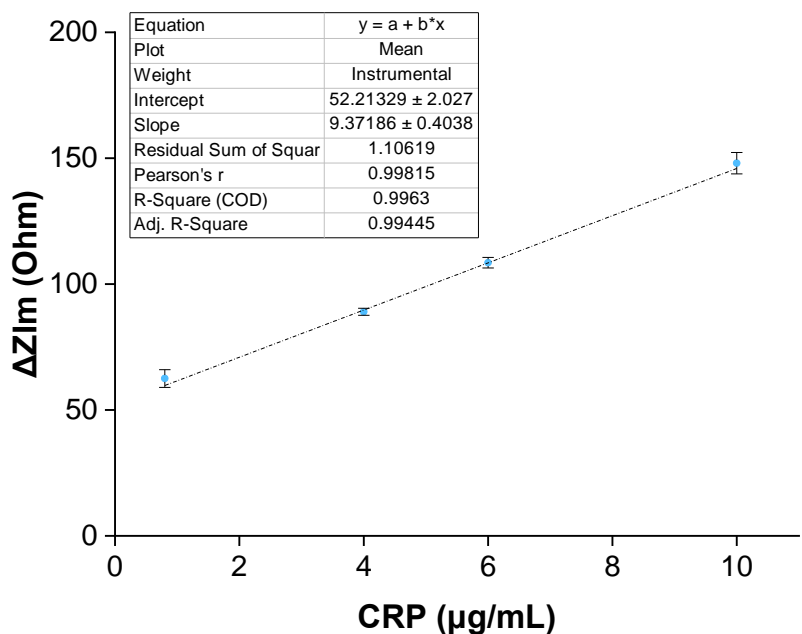

a

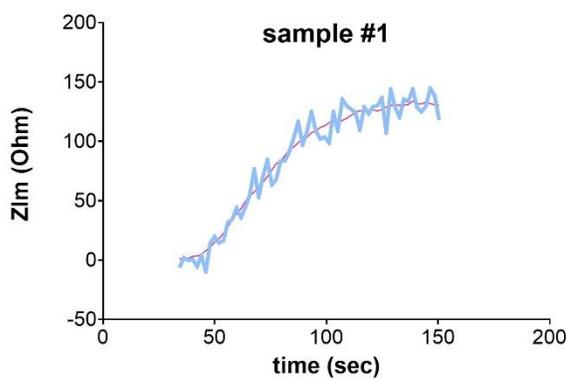

b

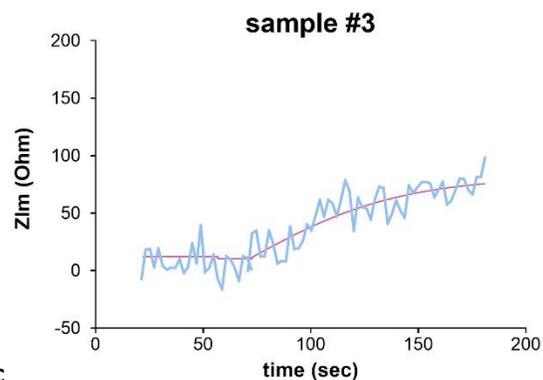

c

Table 1.

| Sample # | Spiked CRP concentration | EIS-Measured CRP concentration | Error (%) |
|---|---|---|---|
| 1 | 7 µg/mL | 7.54 µg/mL | 7.7 |
| 3 | 3 µg/mL | 3.1 µg/mL | 3.4 |

**Figure 5. a)** Calibration curve of the shift of the imaginary part of electrochemical impedance versus CRP concentrations ranging from 0.8 to 10 µg/mL; **b)** and **c):** Blind test curves and their fittings (pink lines) portraying the real-time sensor response upon detection of a) sample #1 and b) sample #3 in ISF-like medium; **Table 1:** Table citing the results of the blind test: the measured CRP concentration, the actual spiked CRP concentration in the blind samples and the calculated measurement error.

## CONCLUSION

In this work, CRP detection was achieved in ISF-like medium using a real-time EIS electrochemical biosensor.

The instrumentation used for CRP detection in the ISF-like solution was greatly compact with a sensing chip of 2 mm$^2$ and a 3D printed flow cell of 2 µL. These dimensions allowed for a low sample consumption of 5 µL. The limit of detection reached was 0.8 µg/mL in the complex ISF-like medium. The response was very fast, recorded for about 100 sec (less than 2 minutes) after the first contact of the CRP sample with the sensing surface, which is promising for future near real-time applications of this type of sensor. Based on the ISF protein concentration, which is 2-3 times lower than that of blood plasma, we argue here, that the sensor developed serves the clinically relevant range of CRP blood tests, especially if one weighs in the facts that the proposed detection needs neither any pre-sample dilution, nor any signal enhancement or labelling via the use of particles or secondary antibodies.

Additionally, the bioassay here, is stable and robust in the inherent complexity of the biofluid exploited. Nevertheless, it owns the ability to achieve lower limit of detection and higher signal-to-noise ratio; sensing electrode optimization, and fit for purpose instrumentation (*e.g.* enhanced electronic readout that will employ advanced filters and amplifiers, and shielding of the electrical connections that will protect them from the environment fluctuations) are expected to enhance the performance of the sensor.

In conclusion, the CRP real-time sensing method presented in this work, is simple, fast, label-free, sensitive, reproducible and low-cost; all these characteristics point to the great potential of this technique to be integrated into a wearable sensing device for ISF biomarkers detection.

## AKNOWLEDGEMENTS

The authors wish to thank Nicolas Ambrosio (Laboratoires, La Source; Lausanne, Switzerland) for conducting the quantification of the CRP samples.